 \documentclass[preprint,onecolumn,amsmath,amssymb,aps,prb]{revtex4-1}

\usepackage{graphicx}% Include figure files
\usepackage{dcolumn}% Align table columns on decimal point
\usepackage{bm}% bold math
 \usepackage{float} 
\usepackage{fullpage}
\usepackage{color}
\usepackage{subcaption}
\usepackage{multirow}
\usepackage{amsmath}
\usepackage{amssymb}
\usepackage{amsthm}
\usepackage[colorlinks   = true, urlcolor     = cyan, linkcolor    = blue, citecolor   = red]{hyperref}
\usepackage{bm}
\usepackage{pict2e}
\usepackage{epstopdf}
\usepackage{comment}
\usepackage{epsfig}
\usepackage{url}
\usepackage[ruled,vlined]{algorithm2e}

\usepackage{cancel}
\usepackage{amsmath}
\usepackage{amssymb}
\usepackage{etoolbox}
\usepackage{graphicx}
 \usepackage[english]{babel}
\usepackage{mathrsfs} % Bilinear forms
\usepackage{enumitem}                                  % Fancy enumerate, e.g. a) b) c)

\usepackage{amsthm} % Theorems
\usepackage{siunitx}
\usepackage{ upgreek }

\begin{document}

%\preprint{APS/123-QED}

\title{Strain engineering of Zeeman and Rashba effects in   transition metal dichalcogenide nanotubes and their Janus variants: An ab initio study}

\author{Arpit Bhardwaj}
\affiliation{College of Engineering, Georgia Institute of Technology, Atlanta, GA 30332, USA}

\author{Phanish Suryanarayana}
\email{phanish.suryanarayana@ce.gatech.edu}
%\homepage{https://www.phanishgroup.com}
%\email{phanish.suryanarayana@ce.gatech.edu}
\affiliation{College of Engineering, Georgia Institute of Technology, Atlanta, GA 30332, USA}

%\date{\today}

\begin{abstract}
We study the influence of mechanical deformations on the Zeeman and Rashba effects in  synthesized transition metal dichalcogenide (TMD)  nanotubes and their Janus variants from first principles. In particular,   we perform symmetry-adapted density functional theory simulations with spin-orbit coupling to determine the variation in the Zeeman and Rashba splittings with axial and torsional deformations. We find significant splitting in molybdenum and tungsten nanotubes, for which the Zeeman splitting decreases with increase in strain, going to zero for large enough tensile/shear strains, while the Rashba splitting coefficient increases linearly with shear strain, while being zero for all tensile strains, a consequence  of the inversion symmetry remaining unbroken. In addition, the Zeeman splitting is relatively unaffected by nanotube diameter, whereas the Rashba coefficient decreases with increase in diameter.  Overall, mechanical deformations represent a powerful tool for spintronics in TMD  nanotubes as well as their Janus variants.
\end{abstract}


\maketitle

%%%%%%%%%%%%%%%%%%%%%%%%%%%%%%%%%%%%%%%%%%%%%%%%%%%%%%%%%%%%%%
\section{Introduction}
Transition metal dichalcogenide (TMD) nanotubes are 1D materials of the  form MX\textsubscript{2}, where  M and  X represent a transition metal and chalcogen, respectively \cite{rao2003inorganic}. They represent the most diverse group of nanotubes, there being 38 transition metals and 3 chalcogens, resulting in a total of 114 possible combinations. Of these,  around 12 have already been synthesized, which represents a significant fraction of the total number of experimentally realized nanotubes, and the most in any group  \cite{tenne2003advances, rao2003inorganic, serra2019overview}.  The number of such nanotubes doubles when  considering their Janus variants \cite{yagmurcukardes2020quantum} --- nanotubes of the form MXY, where Y represents a chalcogen that is distinct from X ---  of which WSSe has  recently been synthesized \cite{sreedhara2022nanotubes}.

TMD nanotubes and their Janus variants demonstrate varying electronic properties, ranging from semiconducting \cite{seifert2000structure, seifert2000electronic, mikkelsen2021band, tao2018band, bhardwaj2022strain} to  metallic \cite{seifert2000novel, enyashin2005computational, bhardwaj2022strain} to  superconducting \cite{nath2003superconducting, tsuneta2003formation}. Notably, these properties can be tuned/engineered by a number of mechanisms, including mechanical deformation \cite{zibouche2014electromechanical,  li2014strain, wang2016strain, lu2012strain, levi2015nanotube, oshima2020geometrical, bhardwaj2022strain}, electric field \cite{wang2014tuning, zibouche2019strong}, temperature \cite{nath2003superconducting, tsuneta2003formation}, chirality/radius \cite{ivanovskaya2003computational, gao2017structural, yin2016chiral, zibouche2012layers, seifert2000electronic, ansari2015ab}, and defects \cite{tal2001effect, li2015tailoring}. This makes the nanotubes  ideally suited for various technological applications, including mechanical sensors \cite{li2016low, sorkin2014nanoscale, oshima2020geometrical}, nanoelectromechanical (NEMS) devices \cite{yudilevichself, levi2015nanotube, divon2017torsional}, biosensors \cite{barua2017nanostructured}, photodetectors \cite{unalan2008zno, zhang2012high, zhang2019enhanced, tang2018janus, oshima2020geometrical, xie2021theoretical, ju2021tuning, ju2021rolling, zhang2019mosse}, and superconductive materials \cite{nath2003superconducting, tsuneta2003formation}. However, the potential for TMD and Janus TMD nanotubes (and nanotubes in general)  to be used in spintronic applications has not been studied  heretofore, particularly in the context of first principles calculations. 

Spintronics or spin electronics refers to the exploitation of both spin  and the electronic charge in solid state devices \cite{cheng2013spin}. In this context,  the Zeeman and Rashba effects are of particular interest, both being relativistic effects arising from spin-orbit coupling (SOC). In particular, the Zeeman and Rashba effects result in  splitting of the electronic bands along the energy and wavevector axes, respectively, of particular importance being those at the valence band maximum (VBM) and the conduction band minimum (CBM). These effects have been studied in TMD monolayers and their Janus variants not only experimentally \cite{larentis2018large, li2020enhanced, jiang2017zeeman}, but also theoretically  using ab initio Kohn-Sham density functional theory (DFT) calculations \cite{cheng2013spin, rezavand2021stacking, chen2020tunable}. In addition, the effect of strain on the Zeeman and Rashba splittings has been studied in Janus TMD bilayers \cite{rezavand2021stacking} and their heterostructures \cite{rezavand2022tuning} using DFT. However, there have been no such studies for TMD and Janus TMD nanotubes (and nanotubes in general), which provides the motivation for the current investigation. 

In this work, we study the influence of mechanical deformations on the Zeeman and Rashba effects in the synthesized TMD nanotubes and their Janus variants using Kohn-Sham DFT calculations. In particular,  we perform symmetry-adapted DFT simulations with SOC to determine the variation in the Zeeman and Rashba splittings with axial and torsional deformations. We find significant splitting for the nanotubes having the transition metal as either molybdenum or tungsten. In particular, axial and torsional deformations can be used to vary the Zeeman splitting, while torsional deformations can be used to introduce and vary the Rashba splitting, making the nanotubes particularly well-suited for spintronics applications. 

The remainder of this manuscript is organized as follows. In Section~\ref{Sec:Methods}, we list the standard and Janus TMD nanotubes studied and describe the symmetry-adapted Kohn-Sham DFT simulations for the calculation of the Zeeman and Rashba splittings.  Next, we  present and discuss the results of the simulations  in Section~\ref{Sec:Results}. Finally, we provide concluding remarks in Section~\ref{Sec:Conclusions}.

%%%%%%%%%%%%%%%%%%%%%%%%%%%%%%%
%%%%%%%%%%%%%%%%%%%%%%%%%%%%%%%

\section{Systems and methods} \label{Sec:Methods}
We start by considering the TMD nanotubes that have been synthesized \cite{tenne2003advances, rao2003inorganic, serra2019overview}: \{MoS\textsubscript{2}, MoSe\textsubscript{2}, MoTe\textsubscript{2}, WS\textsubscript{2}, WSe\textsubscript{2}, WTe\textsubscript{2}, NbS\textsubscript{2}, NbSe\textsubscript{2}, TaS\textsubscript{2}, TiS\textsubscript{2}, TiSe\textsubscript{2}, HfS\textsubscript{2}, and  ZrS\textsubscript{2}\}. Since we have found that spin-orbit coupling (SOC) does not cause any splitting in the nanotubes: \{NbS\textsubscript{2}, NbSe\textsubscript{2}, TaS\textsubscript{2}, TiS\textsubscript{2}, TiSe\textsubscript{2}, HfS\textsubscript{2}, and  ZrS\textsubscript{2}\}, we  henceforth consider the remaining TMD nanotubes: \{MoS\textsubscript{2}, MoSe\textsubscript{2}, MoTe\textsubscript{2}, WS\textsubscript{2}, WSe\textsubscript{2}, WTe\textsubscript{2}\},  as well as their Janus variants with the heavier chalcogen on the outside: \{MoSSe,  MoSTe, MoSeTe, WSSe, WSTe, WSeTe\}, all with 2H-t symmetry. We consider their armchair configurations, since the results remain unchanged for the zigzag configuration, in agreement with previous observations for SOC in the MoS\textsubscript{2} nanotube \cite{milivojevic2020spin}. The diameters of the TMD nanotubes are chosen to be commensurate with those synthesized, and the diameters of the Janus TMD nanotubes are set to DFT-calculated equilibrium values  (Table~\ref{Tab:Diameters}). The axial and torsional deformations considered are also commensurate with those in experiments \cite{levi2015nanotube, divon2017torsional, nagapriya2008torsional, kaplan2007mechanical, kaplan2006mechanical}. Indeed, through phonon calculations using ABINIT \cite{gonze2002first},  we have verified that the  monolayer counterparts are stable at the largest tensile/shear strains (Supplementary Material), which suggests the stability of the nanotubes at the chosen strains, as curvature effects on the phonon spectrum are  expected to be minor at the relatively large diameters of the nanotubes.

 \begin{table}[h!]
		\centering
\caption{Diameters of the TMD nanotubes  \cite{bhardwaj2021torsional, bhardwaj2021strain} and their Janus variants \cite{bhardwaj2021elastic, bhardwaj2022strain}.}
		\resizebox{0.90\textwidth}{!}{
			\begin{tabular}{|c|c|c|c|c|c|}
\hline
	{Material}&{Diameter (nm)}&{Material}&{Diameter (nm)}&{Material}&{Diameter (nm)}\\
				\hline
       {MoS$_2$} & {3.2}  & {MoSe$_2$} & {3.2}& {MoTe$_2$} & {3.6}\\
            \hline

    {WS$_2$} & {3.2} & {WSe$_2$} & {3.2}& {WTe$_2$} & {3.5}\\ \hline
     {MoSSe} & {8.4} & {MoSTe} & {3.8} & {MoSeTe} & {6.6}\\
            \hline
    {WSSe} & {8.8} & {WSTe} & {3.8} &{WSeTe} & {6.6}  \\
            \hline
 
    \end{tabular}}
 \label{Tab:Diameters}   
    \end{table}

\begin{figure}[htbp!]
\centering
\includegraphics[width=0.6\textwidth]{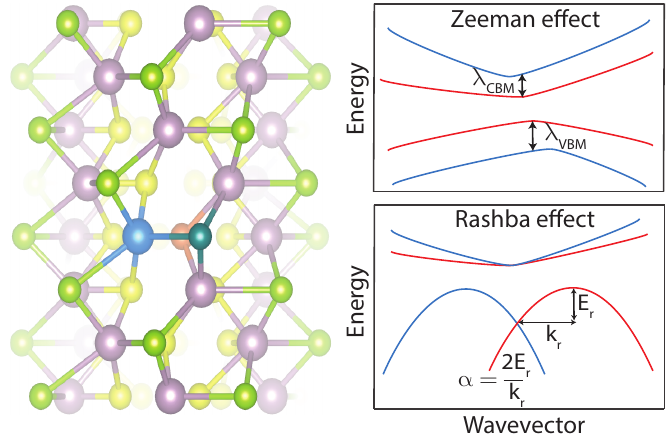}
\caption{Left: Illustration of a Janus TMD nanotube subject to both axial and  torsional deformations, with the three fundamental domain atoms colored blue, orange, and dark green (structural model generated using VESTA \cite{momma2008vesta}). Right: Illustration of the Zeeman and Rashba splittings in the electronic band structure, with $\lambda_{\rm VBM}$ and  $\lambda_{\rm CBM}$ denoting the Zeeman split at the valence band maximum (VBM) and conduction band minimum (CBM), respectively, and $\alpha$ representing the Rashba splitting coefficient.}
\label{fig:illustration}
\end{figure}

We perform Kohn-Sham DFT simulations using the Cyclix-DFT \cite{sharma2021real} feature in the state-of-the-art real-space code SPARC \cite{xu2021sparc, zhang2023versionS, ghosh2017sparc1}. In particular, we perform symmetry-adapted calculations that exploit the cyclic and/or helical symmetry in the system to reduce the Kohn-Sham problem to  the unit cell/fundamental domain  with minimal number of atoms \cite{sharma2021real, ghosh2019symmetry}, e.g., the fundamental domain for the chosen nanotubes contains only 3 atoms,  i.e., 1 metal and 2 chalcogen atoms (Fig.~\ref{fig:illustration}). This reduction due to symmetry can be exploited even on the application of axial and/or torsional deformations, tremendously lowering the computational expense,  given that DFT calculations scale cubically with system size, making otherwise impractical calculations routine, e.g., a $8.5$ nm diameter MoSSe  nanotube with an external twist of $6\times10\textsuperscript{-4}$ rad/bohr has $219,888$ atoms in the simulation domain when employing periodic boundary conditions, a system size that is impractical even with  state-of-the-art approaches \cite{gavini2022roadmap}. Cyclix-DFT is now a mature open source  feature in SPARC, having been verified by comparisons with established DFT codes \cite{sharma2021real}, and ability to make accurate predictions in diverse physical applications \cite{codony2021transversal, kumar2021flexoelectricity, kumar2020bending, bhardwaj2021elastic, bhardwaj2021torsional, bhardwaj2021strain, bhardwaj2022strain, bhardwaj2023ab}. 

In all simulations, we employ the Perdew–Burke–Ernzerhof (PBE) \cite{perdew1996generalized} exchange-correlation functional, and  optimized norm-conserving Vanderbilt (ONCV) \cite{hamann2013optimized} pseudopotentials with nonlinear core correction (NLCC) and SOC from the PseudoDojo  collection \cite{van2018pseudodojo}. The equilibrium geometry of the nanotubes (Supplementary Material) is in very good agreement with previous DFT calculations \cite{li2014strain, chang2013orbital, luo2019electronic, mikkelsen2021band, wang2018mechanical, bolle2021structural}, and the equilibrium geometry of the corresponding monolayers (Supplementary Material) is in very good agreement with experiments \cite{klots2014probing, ugeda2014giant, hill2016band, lu2017janus} as well as DFT calculations \cite{wang2016strain, li2014strain, haastrup2018computational, shi2018mechanical, zhao2015ultra}, verifying the accuracy of the chosen pseudopotential and exchange-correlation functional. Though more advanced and expensive exchange-correlation functionals such as hybrid generally provide better spectral properties, this is not always the case, e.g., Janus TMD monolayers \cite{bhardwaj2021elastic}, motivating the choice of PBE exchange-correlation here, as done in previous works for such systems \cite{chang2013orbital, mikkelsen2021band, wang2018mechanical, bolle2021structural, shi2018mechanical}. The numerical parameters in the Cyclix-DFT simulations, including grid spacing for real-space discretization, grid spacing for Brillouin zone integration, vacuum in the radial direction, and structural relaxation tolerances, are chosen such that the Zeeman splitting values and Rashba coefficients are converged to within  0.01 eV and 0.01 eV \AA, respectively. This translates to an accuracy of $10^{-4}$ Ha/atom in the ground state energy.

%%%%%%%%%%%%%%%%%%%%%%%%%%%%%%%%%%%%%%%%%%%%%%%%%%%%%%%%%%%%%%%%%%%%%%%%%%%%%%%%%%%%%%%%%%%%%%%%%%%%%%%%%%%%%%%%%%%%%%%%%%%%%%%%%%%%%%%%%%%%%%%%
\section{Results and discussion} \label{Sec:Results}

We now use the aforedescribed framework to study the effect of axial and torsional deformations on the Zeeman and Rashba splittings in the molybdenum and tungsten TMD nanotubes and their Janus variants. In the results presented here, the axial strain ($\varepsilon$) is defined as the change in nanotube length divided by its original length; the shear strain ($\gamma$) is defined as the product of the nanotube radius with the applied twist per unit length; the Zeeman splitting ($\lambda_{\rm VBM}$) corresponds to the valence band maximum (VBM), where the effect is significantly more pronounced ($\sim$5x larger) than the conduction band minimum (CBM) (Supplementary Material); and the Rashba splitting coefficient ($\alpha$) corresponds to the VBM, calculated at the zero wavevector in the axial direction. All the data can be found in the Supplementary Material. 

In Fig.~\ref{fig:Zeeman}, we present the variation in the Zeeman splitting  with tensile and shear strains. We observe that the Zeeman splitting  in the undeformed state is significant, being comparable to the monolayer counterparts \cite{cheng2013spin}, with the WTe\textsubscript{2} and MoS\textsubscript{2}/MoSTe nanotubes having the  largest and smallest values of $\lambda_{\rm VBM} =489$ and $\sim$146 meV, respectively.  In addition, the splitting decreases with increase in tensile strains, going to zero for large enough strains. A similar behavior is also observed for shear strains, other than for the MoSe\textsubscript{2}, WS\textsubscript{2}, WSSe, and WSTe nanotubes, where the splitting remains unaffected by the torsional deformations. Such a decrease in the Zeeman splitting values upon the application of biaxial strains has been observed for Janus TMD bilayers \cite{rezavand2021stacking}. Note that the values for the  Janus TMD nanotubes are generally in between their parent TMDs. Note also that the sudden jumps in the Zeeman splitting values are a consequence of the VBM location shifting to a different wavevector. 

\begin{figure}[htbp!]
    \centering
    \includegraphics[width=0.9\textwidth]{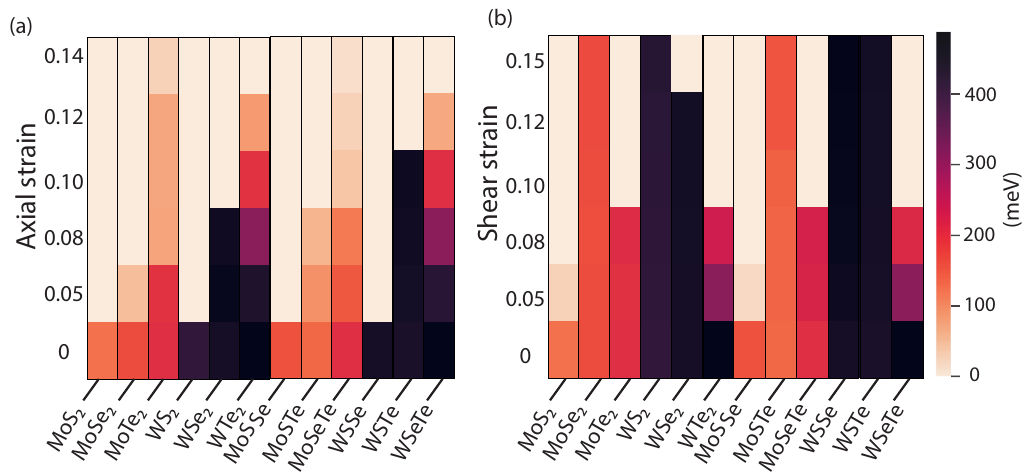}
    \caption{Variation in the Zeeman splitting ($\lambda_{\rm VBM}$) for the molybdenum and tungsten TMD nanotubes and their Janus variants  with (a) tensile strain ($\varepsilon$) and (b) shear strain ($\gamma$). The tensile and shear strains result from the application of axial and torsional deformations, respectively.}
      \label{fig:Zeeman}
    \end{figure}

%%%%%%%%%%

\begin{figure}[htbp!]
    \centering
    \includegraphics[width=0.49\textwidth]{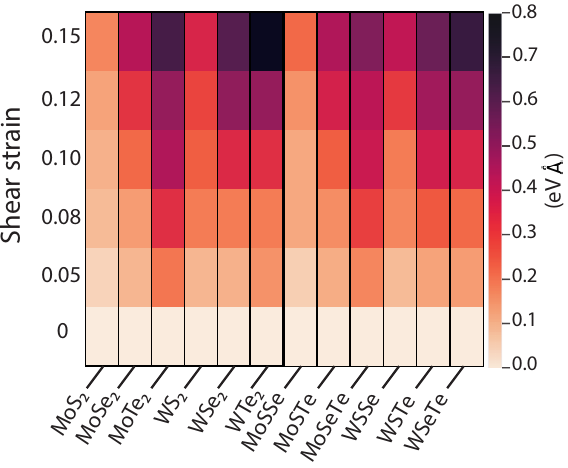}
    \caption{Variation in the Rashba splitting coefficient ($\alpha$)  for the molybdenum and tungsten TMD nanotubes and their Janus variants  with shear strain ($\gamma$). The shear strain results from the application of torsional deformations.}
     \label{fig:Rashba}
\end{figure}

In Fig~\ref{fig:Rashba}, we present the variation in the Rashba  coefficient  with shear strain.  Unlike torsional deformations, axial deformations do not break the inversion symmetry of the nanotube, and therefore the  Rashba effect remains absent \cite{naaman2012chiral}. We observe that the Rashba coefficient  increases linearly with shear strain --- average coefficient of determination of linear regression over all the materials is 0.97 --- reaching significantly large values that are comparable to those for Janus TMD monolayers \cite{rezavand2021stacking}, systems where we have found the Rashba effect to be  insensitive to shear strains. Indeed, the Rashba effect is not observed in TMD monolayers due to the presence of inversion symmetry. At the largest shear strain of $\gamma = 0.15$, the largest and smallest Rashba coefficient values of $\alpha = 0.78$ and 0.20 eV \AA \, occur for the WTe\textsubscript{2} and MoS\textsubscript{2} nanotubes, respectively, whose undeformed configurations  also have the largest and smallest Zeeman splitting, respectively. However, this correlation is not generally true, e.g., MoTe\textsubscript{2} has one of the smallest Zeeman splitting of $\lambda_{\rm VBM} = 213$ meV for the undeformed nanotube, whereas it has one of the largest Rashba coefficient of $\alpha = 0.65$   eV \AA \, for the maximum shear strained tube ($\gamma=0.15$). Note that the values for the  Janus TMD nanotubes are generally in between their parent TMDs.

%%%%%%%%%%

\begin{figure}[htbp!]
        \centering
        \includegraphics[width=0.65\textwidth]{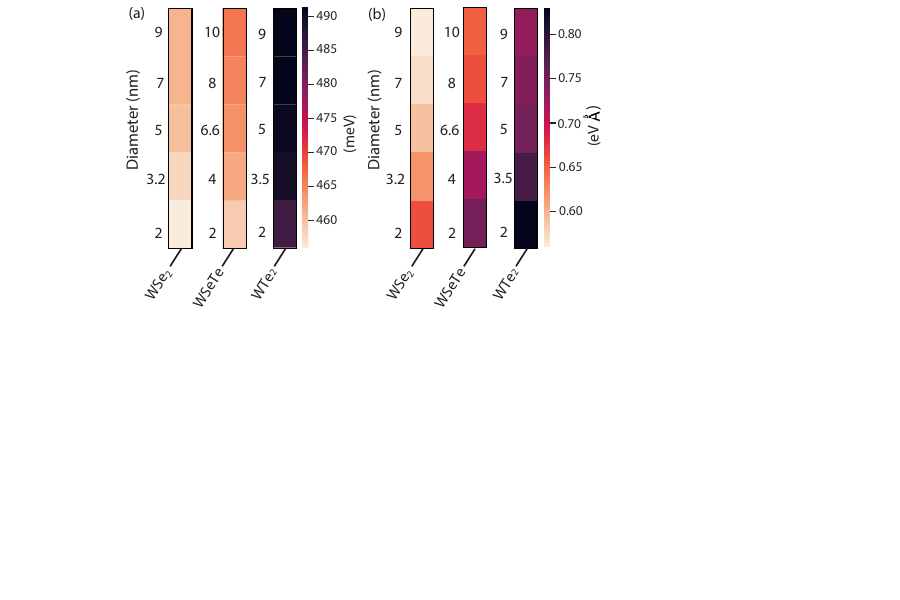}
        \caption{Variation in the (a) Zeeman splitting  ($\lambda_{\rm VBM}$) and (b) Rashba coefficient ($\alpha$) with diameter.  The Zeeman splitting corresponds to undeformed nanotube, while the Rashba coefficient corresponds to the maximum shear strained ($\gamma=0.15$) nanotube.}
      \label{fig:Curvature}
    \end{figure}

To understand the effect of the nanotube diameter on the results obtained, we now consider the nanotubes that demonstrate the largest Zeeman and Rashba effects, i.e., WSe\textsubscript{2}, WSeTe, and WTe\textsubscript{2}, with diameters spanning the range $\sim 2 - 10$ nm. In Fig.~\ref{fig:Curvature}, we present the variation in the Zeeman splitting  and Rashba coefficient  with the diameter, while considering the unstrained and largest shear strain ($\gamma = 0.15$) configurations, respectively. We observe that the Zeeman splitting values remain relatively unchanged, increasing ever so slightly with diameter --- around 1\% over the entire diameter range --- approaching the flat sheet values  of $\lambda_{\rm VBM} = 463$, 473, and 493 meV for  WSe\textsubscript{2}, WSeTe, and WTe\textsubscript{2}, respectively \cite{cheng2013spin}. In addition, the Rashba coefficient decreases significantly with increase in diameter, e.g., the value for WTe\textsubscript{2} reduces from $\alpha = 0.83$ eV \AA \, at a diameter of 2 nm to $\alpha = 0.73$ eV \AA \, at a diameter of  9 nm, expectedly heading towards the  zero value  for the flat sheet configuration.

The  results presented here clearly demonstrate that mechanical deformations can be used to engineer the Zeeman and Rashba splittings in molybdenum and tungsten TMD nanotubes as well as their Janus variants, making them a powerful tool for spintronics applications. In particular, the Zeeman effect is especially significant for the undeformed nanotubes, becoming progressively smaller and even disappearing with increase in axial/shear strains, and  the Rashba effect can be introduced through torsional deformations --- break the inversion symmetry of the system --- becoming especially significant as the shear strain increases.

%%%%%%%%%%%%%%%%%%%%%%%%%%%%%%%%%%%%%%%%%%%%%%
%%%%%%%%%%%%%%%%%%%%%%%%%%%%%%%%%%%%%%%%%%%%%%
%%%%%%%%%%%%%%%%%%%%%%%%%%%%%%%%%%%%%%%%%%%%%%

\section{Concluding remarks} \label{Sec:Conclusions}
In this work, we have studied the strain engineering of Zeeman and Rashba effects in  synthesized TMD nanotubes  and  their Janus variants  using first principles DFT simulations. In particular, we have performed symmetry-adapted Kohn-Sham calculations with spin-orbit coupling to determine the effect of axial and torsional deformations on the Zeeman and Rashba splittings in the electronic band structure. We have found that there is significant splitting in the  molybdenum and tungsten nanotubes, for which the Zeeman splitting decreases with increase in tensile/shear strain, reaching zero for large enough strains, while the Rashba splitting coefficient increases linearly with shear strain, while being zero for all axial deformations, a consequence  of the inversion symmetry remaining unbroken. In addition, the Zeeman splitting is relatively unaffected by the nanotube diameter, whereas the  Rashba coefficient decreases with increase in diameter. Though the current study has been restricted to  TMD nanotubes and their Janus variants, other nanotubes are expected to demonstrate similar behavior, particularly those with heavy chemical elements. Overall, mechanical deformations represent a powerful tool for spintronics applications using nanotubes.

%\ack
%The authors gratefully acknowledge the support of the US National Science Foundation (CAREER-1553212 and MRI-1828187).

\section*{Acknowledgements} 
The authors gratefully acknowledge the support of the Clifford and William Greene, Jr. Professorship. This research was also supported by the supercomputing infrastructure provided by Partnership for an Advanced Computing Environment (PACE) through its Hive (U.S. National Science Foundation (NSF) through grant MRI-1828187) and Phoenix clusters at Georgia Institute of Technology, Atlanta, Georgia.  \vspace{-1mm}

%\bibliographystyle{unsrt}
%\bibliographystyle{iopart-num}
%\bibliography{refer.bib} 

\end{document}